\documentclass[twocolumn,superscriptaddress,showpacs]{revtex4}

 \usepackage{graphicx}

\usepackage{bm,color}

\newcommand {\be}{\begin{equation}}
\newcommand {\ee}{\end{equation}}

\begin{document}

\title{Out-of-equilibrium phase transitions in the HMF model: a closer look}
\author{F. Staniscia}
\affiliation{Dipartimento di Fisica, Universit\`a di Trieste, 34127 Trieste, Italy}
\affiliation{Sincrotrone Trieste, S.S. 14 km 163.5, Basovizza, 34149 Trieste, Italy}
\author{P.H. Chavanis}
\affiliation{Laboratoire de Physique Th\'eorique (IRSAMC), Universit\'e de Toulouse (UPS) and CNRS, F-31062 Toulouse, France}
\author{G. De Ninno}
\affiliation{Sincrotrone Trieste, S.S. 14 km 163.5, Basovizza, 34149 Trieste, Italy}
\affiliation{Physics Department, Nova Gorica University 5001 Nova Gorica, Slovenia}


\begin{abstract}

We provide a detailed discussion of out-of-equilibrium phase
transitions in the Hamiltonian Mean Field (HMF) model in the framework
of Lynden-Bell's statistical theory of the Vlasov equation. For
two-levels initial conditions, the caloric curve $\beta(E)$ only
depends on the initial value $f_0$ of the distribution function. We
evidence different regions in the parameter space where the nature of
phase transitions between magnetized and non-magnetized states
changes: (i) for $f_0>0.10965$, the system displays a
second order phase transition; (ii) for
$0.109497<f_0<0.10965$, the system
displays a second order phase transition and a first order phase
transition; (iii) for
$0.10947<f_0<0.109497$, the system
displays two second order phase transitions; (iv) for
$f_0<0.10947$, there is no phase transition. The
passage from a first order to a second order phase transition
corresponds to a tricritical point. The sudden appearance of two
second order phase transitions from nothing corresponds to a second
order azeotropy.  This is associated with a phenomenon of phase
reentrance.  When metastable states are taken into account, the
problem becomes even richer. In particular, we find a new situation of
phase reentrance. We consider both microcanonical and canonical
ensembles and report the existence of a tiny region of ensembles
inequivalence. We also explain why the use of the initial
magnetization $M_0$ as an external parameter, instead of the phase
level $f_0$, may lead to inconsistencies in the thermodynamical
analysis.
\end{abstract}



\maketitle


\section{Introduction}
\label{sec_intro}

Systems with long-range interactions have recently been the object of
an intense activity \cite{houches,assise,oxford,cdr}. These systems
are numerous in nature and concern different disciplines such as
astrophysics (galaxies) \cite{paddy,katz,ijmpb}, two-dimensional
turbulence (vortices) \cite{tabeling,sommeria,houchesPHC}, biology
(chemotaxis) \cite{murray}, plasma physics
\cite{nicholson,davidson,dubin} and modern technologies such as Free
Electron Lasers (FEL) \cite{generalref1,generalref2,generalref3}.  In
addition, their study is interesting at a conceptual level because it
obliges to go back to the foundations of statistical mechanics and
kinetic theory \cite{assisePHC,cdr,bgm,kinuni}. Indeed, systems with
long-range interactions exhibit a number of unusual features that are
not present in systems with short-range interactions. For example,
their equilibrium statistical mechanics is marked by the existence of
spatially inhomogeneous equilibrium states \cite{houches}, unusual
thermodynamic limits \cite{messer,kiessling,caglioti}, inequivalence
of statistical ensembles \cite{paddy,ijmpb,ellis}, negative specific
heats \cite{thirring,lbheat}, various kinds of phase transitions
\cite{ijmpb,bb} etc. Their dynamics is also very interesting because
these systems can be found in long-lived quasi stationary states (QSS)
that are different from Boltzmann equilibrium states. These QSSs can
be interpreted as stable steady states of the Vlasov equation which
governs the evolution of the system for sufficiently ``short'' times
before correlations have developed \cite{nicholson,bt}. In fact, for
systems with long-range interactions, the collisional relaxation time
towards the Boltzmann distribution increases rapidly with the number
of particles $N$ and diverges at the thermodynamic limit $N\rightarrow
+\infty$ \cite{bt,cdr,bgm,kinuni}. Therefore, the domain of validity
of the Vlasov equation is huge and the QSSs have very long
lifetimes. In many cases, they are the only observable structures in a
long-range system, so that they are often more physically relevant
than the Boltzmann equilibrium state itself. A question that naturally
emerges is whether one can predict the QSS actually reached by the
system. This is not an easy task since the Vlasov equation admits an
infinite number of stable steady states in which the system can be
trapped \cite{bt}. In a seminal paper, Lynden-Bell \cite{lb} proposed
to determine the QSS eventually reached by the system by developing a
statistical mechanics of the Vlasov equation. To that purpose, he
introduced the notions of phase mixing, violent relaxation and
coarse-grained distributions. He obtained the most probable
distribution by maximizing a Boltzmann-type entropy while conserving
all the constraints of the Vlasov equation (in particular the infinite
class of Casimirs). By definition, this ``most mixed state'' is the
statistical equilibrium state of the Vlasov equation (at a
coarse-grained scale). Whether or not the system truly reaches this
equilibrium state relies on an assumption of ergodicity and efficient
mixing. This ergodicity assumption is not always fulfilled in the
process of violent relaxation and the Lynden-Bell prediction may
fail. In that case, the QSS can be another stable steady state of the
Vlasov equation that is incompletely mixed. This is referred to as
incomplete relaxation (see, e.g. \cite{incomplete}, for discussion and
further references). In case of incomplete relaxation, the prediction
of the QSS is very difficult, and presumably impossible. Nevertheless,
in many cases, the Lynden-Bell approach gives a fine first order
prediction of the achieved QSS and allows one to predict
out-of-equilibrium phase transitions between different types of
structures that can be compared with direct simulations or
experiments. Before addressing this problem in a specific situation,
namely the Hamiltonian Mean Field (HMF) model \cite{inagaki,ar}, let
us first briefly review the successes and the weaknesses of the
Lynden-Bell approach.

Lynden-Bell's statistical theory of violent relaxation was elaborated
in the context of 3D stellar systems. Unfortunately, this is the worse
situation for its practical application. Indeed, the predicted
distribution function has infinite mass (the spatial density decreases
at large distances like $r^{-2}$). In other words, this means that
there is no entropy maximum for a stellar system in an infinite domain
\cite{bt}. This is a clear evidence of the fact that galaxies have
necessarily reached a state of {\it incomplete} violent relaxation. In
fact, the Lynden-Bell theory is able to explain the isothermal core of
elliptical galaxies without recourse to collisions that operate on a
much longer timescale (of the order of the Chandrasekhar relaxation
time \cite{chandrabook}). This is usually recognized as a major
success of the theory. Unfortunately, it fails at predicting the
structure of the halo whose velocity distribution is anisotropic and
whose spatial density decreases like $r^{-4}$ \cite{bt}. Models of
incomplete violent relaxation have been elaborated by Bertin \&
Stiavelli \cite{bs}, Stiavelli \& Bertin \cite{sb} and Hjorth \&
Madsen \cite{hm}. These models are able to reproduce the de
Vaucouleurs law of elliptical galaxies and provide a very good
agreement with numerical simulations up to nine orders of magnitude
\cite{trenti}. Another possibility to describe incomplete relaxation
is to develop a kinetic theory of violent relaxation in order to
understand what limits mixing \cite{csr,mnraskin,hb4}. The idea is that, in
case of incomplete relaxation (non-ergodicity), the prediction of the
QSS is impossible without considering the dynamics
\cite{incomplete}. Finally, in more academic studies \cite{csmnras},
one can confine the system within an artificial spherical box and
assume a complete relaxation inside the box. Since the Lynden-Bell
distribution is similar to the Fermi-Dirac statistics (in the
two-levels approximation), the problem is mathematically equivalent to
the study of a gas of self-gravitating fermions in a box. This
theoretical problem has been studied in detail by Chavanis
\cite{prefermions}. The caloric curve $\beta(E)$ displays a rich
variety of microcanonical and canonical phase transitions (zeroth and
first order) between gaseous (non degenerate) and condensed
(degenerate) states, depending on the value of a degeneracy parameter
related to the initial distribution function $f_0$ in the Lynden-Bell
theory. In particular, there exists two critical points in the phase
diagram, one in each ensemble, at which the phase transitions are
suppressed. For details about these phase transitions, and for an
extended bibliography, we refer to the review \cite{ijmpb}. The
Lynden-Bell prediction has also been tested in 1D and 2D gravity
\cite{yama,levine2D} where the infinite mass problem does not arise
\cite{fermid}. However, it is found again that relaxation is
incomplete and that the Lynden-Bell prediction fails \footnote{Teles
{\it et al.} \cite{levine2D} propose a variant of the Lynden-Bell
distribution that gives very good agreement with numerical
simulations.}. Finally, Arad \& Lynden-Bell \cite{arad} have shown
that the theory itself presents some inconsistencies arising from its
non-transitive nature. These negative results have led many
astrophysicists to the conclusion that the Lynden-Bell theory does not
work in practice \cite{bt}.

A similar statistical theory has been developed by Miller
\cite{miller}, and independently by Robert \& Sommeria \cite{rs}, in
2D turbulence in order to explain the robustness of long-lived
vortices in astrophysical and geophysical flows (a notorious example
being Jupiter's great red spot). Large-scale vortices are interpreted
as quasi stationary states of the 2D Euler equation in the same way
that galaxies are quasi stationary states of the Vlasov equation (see
\cite{houchesPHC,kindetail} for a discussion of the numerous analogies
between the statistical mechanics and the kinetic theory of 2D
vortices and stellar systems). Miller-Robert-Sommeria (MRS) developed
a statistical theory of the 2D Euler equation in order to predict the
most probable state achieved by the system.  Although situations of
incomplete relaxation have also been evidenced in 2D turbulence
\cite{hd,boghosian,brands}, the MRS theory has met a lot of
success. For example, it is able to account for geometry induced phase
transitions between monopoles and dipoles as we change the aspect
ratio of the domain \cite{jfm1,jfm2,vb,naso,taylor}. Phase transitions and
bifurcations between different types of flows have also been studied
in \cite{staquet,chen,wt}. On the other hand, when applied to
geophysical and astrophysical flows, the MRS theory is able to account
for the structure and the organization of large-scale flows such as
jovian jets and vortices \cite{turk,eht,bsj,sw} and Fofonoff flows in
oceanic basins \cite{vb,ncd}. This theory has also been applied to
more complicated situations such as 2D magnetohydrodynamics (MHD)
\cite{jordan,leprovost} and axisymmetric flows (the celebrated von
K\'arm\'an flow) \cite{karman}.

A toy model of systems with long-range interactions, called the
Hamiltonian Mean Field (HMF) model, has been introduced in statistical
physics \cite{inagaki,ar} and extensively studied \cite{cdr}. It can
be viewed as a $XY$ spin system with infinite range interactions or as
a one dimensional model of particles moving on a ring and interacting
via a long-range potential truncated to one Fourier mode (cosine
potential). In that second interpretation, it shares many analogies
with self-gravitating systems
\cite{inagaki,ar,cvb} but is much simpler to study since it
avoids difficulties linked with the singular nature of the
gravitational potential at the origin and the absence of a natural
confinement \cite{paddy,katz,ijmpb}. The observation of quasi
stationary states in the HMF model
\cite{ar,latora} was a surprise in the community of statistical
mechanics working on systems with long-range interactions. It was
recognized early that these QSSs are out-of-equilibrium structures and
that they are non-Boltzmannian. They were first interpreted
\cite{latora} in terms of Tsallis generalized thermodynamics
\cite{tsallis} with the argument that the system is nonextensive so
that Boltzmann statistical mechanics is not applicable. Later,
inspired by analogies with stellar systems and 2D vortices reported in
\cite{houchesPHC}, different groups started to interpret these QSSs in
terms of stable steady states of the Vlasov equation and statistical
equilibrium states in the sense of Lynden-Bell
\cite{yamaguchi,cvb,barre}.  Chavanis \cite{epjb} studied
out-of-equilibrium phase transitions in the HMF model by analogy with
similar studies in astrophysics and hydrodynamics \cite{csmnras,jfm1}
and obtained a phase diagram in the $(f_0,E)$ plane \footnote{We shall
explain in Sec. \ref{sec_lbt} why these control parameters are the
proper ones to consider in the Lynden-Bell theory.} between magnetized
($M\neq 0$) and non-magnetized ($M=0$) states. These regions are
separated by a critical line $E_c(f_0)$ that marks the domain of
stability of the homogeneous phase. This critical line displays a
turning point at $((f_0)_*,E_*)\simeq (0.10947,0.608)$ leading to a
phenomenon of {\it phase reentrance} (as we reduce the energy, the
homogeneous phase is successively stable, unstable and stable
again). Antoniazzi {\it et al.} \cite{pre} studied the validity of the
Lynden-Bell prediction by performing careful comparisons with direct
$N$-body simulations at $E=0.69$ and found a good agreement for
initial magnetizations $M_0<(M_0)_{crit}(E)\simeq 0.897$ leading to
spatially homogeneous Lynden-Bell distributions
\footnote{Discrepancies occur for larger values of the magnetization,
in particular for $M_0=1$ \cite{ar,latora,campa}. This was interpreted
as a result of incomplete relaxation in \cite{epjb}. Chavanis \& Campa
\cite{cc} showed that polytropic (Tsallis) distributions can provide a
good description of QSSs in some cases of incomplete relaxation. In
their work, Tsallis distributions are justified not by the fact that
the system is nonextensive but by the fact that the evolution is
non-ergodic (i.e. the system does not mix well).}. Antoniazzi {\it et al.} \cite{tri} obtained a phase
diagram in the $(M_0,E)$ plane and showed that the system exhibits
first and second order phase transitions separated by a tricritical
point. Finally, Antoniazzi {\it et al.} \cite{antovlasov} performed
numerical simulations of the Vlasov equation and found a good
agreement with direct $N$-body simulations and Lynden-Bell's
prediction for the explored range of parameters. A synthesis of these
results was published in \cite{proc}.  In this paper, a more detailed
discussion of phase transitions in the $(f_0,E)$ plane was given,
showing the lines of first and second order phase transitions and the
domains of metastability. On the other hand, a comparison between the
phase diagrams in the $(f_0,E)$ and $(M_0,E)$ planes was made. It was
stated, without rigorous justification, that the tricritical point in
the $(M_0,E)$ plane corresponds to the turning point of the critical
line $E_c(f_0)$ in the $(f_0,E)$ plane, i.e. the point where the phase
reentrance starts. These results were confronted to numerical
simulations by Staniscia {\it et al.} \cite{staniscia}. These
simulations confirmed the existence of a reentrant phase in the very
narrow region predicted by the theory \cite{epjb} but also showed
discrepancies with the Lynden-Bell prediction (such as an additional
reentrant phase and a persistence of magnetized states in the {\it a
priori} non-magnetized region) that were interpreted as a result of
incomplete relaxation.  Staniscia {\it et al.}  \cite{staniscia} also
determined the physical caloric curve $\beta_{kin}(E)$, where
$\beta_{kin}=1/T_{kin}$ is the inverse kinetic temperature, in the
region of the phase diagram displaying first and second order phase
transitions, and reported the existence of a region of negative
kinetic specific heat $C_{kin}=dE/dT_{kin}<0$. In a recent paper
\cite{staniscia2}, the thermodynamical caloric curve $\beta(E)$ was
determined in the same region of parameters and it was shown that the
thermodynamical specific heat $C=dE/dT$ is always positive, even in
the region where the kinetic specific heat is negative. In particular,
it is argued that the ensembles are equivalent although the
experimentally measured specific heat is negative \footnote{In
\cite{staniscia2}, it is concluded that it is possible to measure
negative kinetic specific heats in the canonical ensemble when the
distribution function is non-Boltzmannian. In the context of
Lynden-Bell's statistical theory of violent relaxation relying on the
Vlasov equation, the distribution function is non-Boltzmannian
(leading to possibly negative kinetic specific heats) but the only
physically relevant statistical ensemble is the microcanonical one
since the system is isolated (the energy is conserved). If we put the
system in contact with a thermal bath as in \cite{baldovin}, we break
the structure of the Vlasov equation (in the regime where the bath has
some influence on the system's dynamics) and the Lynden-Bell theory
does not apply anymore. In that case, the distribution becomes
Boltzmannian with the temperature of the bath (and the kinetic
specific heat is necessarily positive). In other words, we cannot
impose the temperature of the Lynden-Bell distribution; the relevant
control parameter is the energy.  However, it is suggested in
\cite{staniscia2} that there may exist other situations in which the
system is in contact with a thermal bath imposing its temperature
while the distribution function is non-Maxwellian. The
characterization of this situation is still a matter of
investigation.}.

These various results show that the description of out-of-equilibrium
phase transitions in the HMF model is extremely rich and subtle. In
this paper, we describe in more detail the phase transitions between
magnetized and non-magnetized states in the $(f_0,E)$ plane. In
particular, we plot the series of equilibria $\beta(E)$ for different
values of $f_0$ and determine the caloric curve corresponding to fully
stable states. This completes and illustrates our previous study
\cite{staniscia} where only the final phase diagram was reported. We
evidence different regions in the parameter space where the nature of
phase transitions changes: (i) for $f_0>(f_0)_t\simeq 0.10965$, the
system displays a second order phase transition; (ii) for
$(f_0)_1\simeq 0.109497<f_0<(f_0)_t\simeq 0.10965$, the system
displays a second order phase transition and a first order phase
transition; (iii) for
$(f_0)_*\simeq 0.10947<f_0<(f_0)_1\simeq 0.109497$, the system
displays two second order phase transitions; (iv) for
$f_0<(f_0)_*\simeq 0.10947$, there is no phase transition. The
passage from a first order phase transition to a second order phase
transition corresponds to a tricritical point. The sudden appearance
of two second order phase transitions from nothing corresponds to a
second order azeotropy.  This is associated with a phenomenon of phase
reentrance.  When we take into account metastable states, the
description is even richer and seven regions must be considered (see
Sec. \ref{sec_description}). In particular, we find a new situation of
phase reentrance. We also stress two unexpected results that were not
reported (or incorrectly reported) in previous works: (i) Contrary to
what is stated in \cite{staniscia2}, there exists a region of
ensembles inequivalence but it concerns an extremely narrow range of
parameters so that the conclusions of \cite{staniscia2} are not
altered; (ii) the tricritical point separating second and first order
phase transitions does not exactly coincide with the turning point of
the stability line $E_c(f_0)$, contrary to what is stated in
\cite{proc}, but is slightly different. Again, the difference is small
so that the main results of previous works are not affected. However,
this slight difference leads to an even richer variety of phase
transitions.  We may be fascinated by the fact that so many things
happen in such a very narrow range of parameters (typically
$(f_0)_m\simeq 0.1075<f_0<(f_0)_c\simeq 0.11253954$) although $f_0$ can
take {\it a priori} any value between $0$ and $+\infty$! Finally, we
make clear in this paper (see Sec. \ref{sec_lbt}) that the relevant
control parameters associated with the Lynden-Bell theory are
$(f_0,E)$ \cite{epjb} while the use of the variables $(M_0,E)$
\cite{tri,antovlasov} may lead to physical inconsistencies in the
thermodynamical analysis.

\section{The Lynden-Bell theory and the choice of the control parameters}
\label{sec_lbt}

The HMF model \cite{inagaki,ar}, which shares many similarities with gravitational and charged sheet models, describes the one-dimensional motion of $N$ particles of unit mass moving on a unit circle and  coupled through a mean field cosine interaction. The system Hamiltonian reads
\begin{eqnarray}
\label{lbt1}
H=\frac{1}{2}\sum_{i=1}^{N}v_i^2+\frac{1}{2N}\sum_{i,j=1}^{N}\left \lbrack 1-\cos(\theta_i-\theta_j)\right \rbrack,
\end{eqnarray}
where $\theta_i$ represents the angle that particle $i$ makes with an axis of reference and $v_i$ stands for its velocity. The $1/N$ factor in front of the potential energy corresponds to the Kac prescription to make the system extensive and justify the validity of the  mean field approximation in the limit $N\rightarrow +\infty$. The relevant order parameter is the magnetization defined as ${\bf M}=(\sum_i {\bf m}_i)/N$ where ${\bf m}_i=(\cos\theta_i,\sin\theta_i)$. In the $N\rightarrow +\infty$ limit, the time evolution of the one body distribution function $f(\theta,v,t)$ is governed by the Vlasov equation
\begin{eqnarray}
\label{lbt2}
\frac{\partial f}{\partial t}+v\frac{\partial f}{\partial \theta}-(M_x[f]\sin\theta-M_y[f]\cos\theta)\frac{\partial f}{\partial v}=0,
\end{eqnarray}
where $M_x[f]=\int f(\theta,v,t)\cos\theta\, d\theta dv$ and $M_y[f]=\int f(\theta,v,t)\sin\theta\, d\theta dv$ are the two components of the magnetization.

The statistical theory of the Vlasov equation, introduced by Lynden-Bell \cite{lb}, has been reviewed  in several papers \cite{assisePHC,epjb,pre,cdr,proc,staniscia} so that we shall here only recall the main lines that are important to understand the sequel. We assume that the initial distribution function takes only to values $f(\theta,v,t=0)\in \lbrace 0,f_0\rbrace$. For example, it can be made of one or several patches of uniform distribution  $f(\theta,v,0)=f_0$ surrounded by ``vacuum'' $f(\theta,v,0)=0$. We note that the number and the shape of these patches can be completely arbitrary. For such initial conditions, the quantities conserved by the Vlasov equation are: (i) the value $f_0$ of the initial distribution; (ii) the normalization ${\cal M}=\int f\, d\theta dv=1$; (iii) the energy $E=\frac{1}{2}\int f v^2\, d\theta dv+\frac{1}{2}(1-M^2)$. The fine-grained distribution function $f(\theta,v,t)$  is stirred in phase space but conserves its two values $f_0$ and $0$ at any time, i.e. $f(\theta,v,t)\in \lbrace 0,f_0\rbrace$ $\forall t$. However, as time goes on, the two levels values $f_0$ and $0$ become more and more intermingled as a result of a mixing process (filamentation) in phase space. The coarse-grained distribution $\overline{f}(\theta,v,t)$, which can be viewed as a local average of the fine-grained distribution function, takes values intermediate between $0$ and $f_0$, i.e.  $0\le \overline{f}(\theta,v,t) \le f_0$. It is expected to achieve a steady state $\overline{f}(\theta,v)$ as a result of violent relaxation on a relatively short timescale (a few dynamical times). This corresponds to the QSS observed in the simulations. The most probable, or most mixed state, is obtained by maximizing the Lynden-Bell entropy
\begin{eqnarray}
\label{lbt3}
S=-\int \left\lbrack \frac{\overline{f}}{f_0}\ln \frac{\overline{f}}{f_0}+\left (1-\frac{\overline{f}}{f_0}\right )\ln \left (1-\frac{\overline{f}}{f_0}\right )\right\rbrack\, d\theta dv,\nonumber\\
\end{eqnarray}
while conserving $E$ and ${\cal M}$ (for a given value of $f_0$). This determines the statistical equilibrium state of the Vlasov equation. Note that the whole theory relies on an assumption of ergodicity, i.e. efficient mixing. Our aim here is not to determine the range of validity of the Lynden-Bell theory, so that we shall assume that this assumption is fulfilled (see, e.g. \cite{cc}, for a discussion of incomplete relaxation in the HMF model). We are led therefore to considering the maximization problem
\begin{eqnarray}
\label{lbt4}
\max_{\overline{f}}\left\lbrace S[\overline{f}]\, | \, E[\overline{f}]=E, {\cal M}[\overline{f}]=1\right\rbrace,
\end{eqnarray}
for a given value of $f_0$. The critical points of (\ref{lbt4}), canceling the first order variations of the constrained entropy, are given by the variational principle
\begin{eqnarray}
\label{lbt5}
\delta S-\beta \delta E-\alpha\delta {\cal M}=0,
\end{eqnarray}
where $\beta$ and $\alpha$ are Lagrange multipliers. This yields the Lynden-Bell distribution
\begin{eqnarray}
\label{lbt6}
\overline{f}_{LB}(\theta,v)=\frac{f_0}{1+e^{\beta f_0 \epsilon(\theta,v)+f_0\alpha}},
\end{eqnarray}
where $\epsilon(\theta,v)=v^2/2-M_x[\overline{f}_{LB}]\cos\theta-M_y[\overline{f}_{LB}]\sin\theta$ is the individual energy. In the two-levels approximation, the Lynden-Bell distribution is formally identical to the Fermi-Dirac statistics \cite{lb}. Note that $T=\beta^{-1}=(\partial S/\partial E)^{-1}$ is the {\it thermodynamical temperature}. Since the distribution function (\ref{lbt6}) is non-Boltzmannian, the thermodynamical temperature  differs from the  {\it classical kinetic temperature} $T_{kin}=\int f v^2\, d\theta dv$. This point has been studied specifically in \cite{staniscia2}.

The maximization problem (\ref{lbt4}) corresponds to the microcanonical ensemble (MCE). Since the Lynden-Bell theory is based on the Vlasov equation that describes an isolated system, the microcanonical ensemble is the relevant ensemble to consider (the energy is fixed). We can, however, formally define a canonical ensemble. We introduce the free energy functional $J[\overline{f}]=S[\overline{f}]-\beta E[\overline{f}]$ \footnote{Traditionally, the free energy is defined by $F[f]=E[f]-TS[f]$. However, the Massieu function $J[f]=-\beta F[f]$ turns out to be more convenient in the analysis of phase transitions (see, e.g. \cite{ijmpb}) since it corresponds to the direct Legendre transform of the entropy with respect to the energy. Since $F[f]$ and $J[f]$ are equivalent (recall that $\beta$ is fixed in the canonical ensemble), we shall often refer to $J$ as the ``free energy'' by a slight  abuse of language.} and consider the maximization problem
\begin{eqnarray}
\label{lbt7}
\max_{\overline{f}}\left\lbrace J[\overline{f}]\, | {\cal M}[\overline{f}]=1\right\rbrace,
\end{eqnarray}
for a given value of $f_0$. The maximization problems (\ref{lbt4}) and (\ref{lbt7}) have the same {critical points} since the variational principle
\begin{eqnarray}
\label{lbt8}
\delta J-\alpha\delta {\cal M}=0
\end{eqnarray}
returns Eq. (\ref{lbt5}) (recall that $\beta$ is fixed in the canonical ensemble). In addition, it can be shown at a general level \cite{ellis} that a solution of the canonical problem (\ref{lbt7}) is always a solution of the more constrained dual microcanonical problem (\ref{lbt4}), but that the reciprocal is wrong in case of ensembles inequivalence \footnote{In particular, the thermodynamical specific heat $C=dE/dT$ is necessarily positive in CE while it can be positive or negative in MCE.}. Therefore, even if the canonical ensemble is  not physically justified in the context of Lynden-Bell's theory of violent relaxation, it provides nevertheless a {\it sufficient} condition of microcanonical thermodynamical  stability. It is therefore useful in that respect. In addition, it is interesting on a conceptual point of view to study possible inequivalence between microcanonical and canonical ensembles. Therefore, we shall study in this paper the two maximization problems (\ref{lbt4}) and (\ref{lbt7}), while emphasis and illustrations will be given for the more physical microcanonical case.

Before that, let us recall general notions that will be useful in the sequel (for an extended account, see e.g. \cite{ijmpb}). For a given value of $f_0$, the {\it series of equilibria} is the curve $\beta(E)$ containing all the {\it critical points} of (\ref{lbt4}) or (\ref{lbt7}) (as we have seen, they are the same). The stable part of this curve, in each ensemble, gives the corresponding {\it caloric curve}. In MCE, the control parameter is the energy and the stable states are maxima of entropy $S$ at fixed energy and normalization. This defines the microcanonical caloric curve $\beta(E)$. In CE, the control parameter is the inverse temperature and the stable states are maxima  of free energy $J$  at fixed normalization. This defines the canonical caloric curve $E(\beta)$. The {\it strict caloric curve} contains only {\it fully stable states (S)} that are global entropy maxima at fixed energy and normalization in MCE or global free energy maxima at fixed normalization in CE. The {\it physical caloric curve} contains fully stable and {\it metastable states (M)}, that are local entropy maxima at fixed energy and normalization in MCE or local free energy maxima at fixed normalization in CE. The {\it unstable states (U)}, that are minima or saddle points of the thermodynamical potential, must be rejected. Note that  for systems with long-range interactions, metastable states can have very long lifetimes so that they are very important in practice. By studying the caloric curve $\beta(E)$ for a given value of $f_0$, we can describe {\it phase transitions}. Microcanonical first order phase transition are marked by the discontinuity of the inverse temperature $\beta(E)$ at some energy $E_t$. This corresponds to a discontinuity of the first derivative of entropy $S'(E)=\beta(E)$ at $E_t$ in the energy vs entropy curve. There can exist metastable branches around $E_t$ that possibly end at microcanonical {\it spinodal points}. Microcanonical second order phase transitions are marked by the discontinuity of $\beta'(E)$ at some energy $E_c$. This corresponds to a discontinuity of the second derivative of entropy   $S''(E)=\beta'(E)$ at $E_c$. Similarly, canonical first order phase transitions are marked by the discontinuity of energy $E(\beta)$ at some inverse temperature $\beta_t$. This corresponds to a discontinuity of the first derivative of free energy $J'(E)=-E(\beta)$ at $\beta_t$ in the inverse temperature vs free energy curve. There can exist metastable branches around $\beta_t$ that possibly end at canonical  spinodal points. Canonical second order phase transitions are marked by the discontinuity of $E'(\beta)$ at some inverse temperature $\beta_c$. This corresponds to a discontinuity of the second derivatives of free energy  $J''(\beta)=-E'(\beta)$ at $\beta_c$. Finally, by varying the {\it external parameter} $f_0$, we can describe changes from different kinds of phase transitions at some {\it critical values} of $f_0$ and plot the corresponding {\it phase diagrams} $(f_0,E)$ and $(f_0,\beta)$ in microcanonical and canonical ensembles. This is the programm that we shall follow in this paper.

We emphasize that these general results are valid for the caloric curve $\beta(E)$ where $\beta$ is the inverse thermodynamical temperature, not the inverse kinetic temperature. In particular, the thermodynamical specific heat $C=dE/dT$ is always positive in the canonical ensemble while the kinetic specific heat $C_{kin}=dE/dT_{kin}$ can be positive or negative in the canonical ensemble. This has been illustrated in \cite{staniscia2}.

A last comment is in order. If we consider a  waterbag initial condition in which $f(\theta,v,t=0)=f_0$ in the rectangle $[\-\theta_{min},\theta_{max}]\times [-v_{min},v_{max}]$ and $f(\theta,v,t=0)=0$ outside, it seems convenient to take  as control parameters the initial magnetization $M_0$ and the energy $E$ as done in \cite{tri,antovlasov}. Indeed, the specification of these parameters determines $f_0=\phi(E,M_0)$ and $E$ and thus allows to compute the corresponding Lynden-Bell state. Therefore, it seems that the choice of the control parameters $(E,M_0)$ or $(E,f_0)$ is just a question of commodity. In fact, this is not the case, and we would like to point out some difficulties in taking $(E,M_0)$ as control parameters in the thermodynamical analysis:

(i) The control parameters $(E,M_0)$ are less general than $(E,f_0)$ because they assume that the initial condition is a waterbag distribution, whereas the control parameters $(E,f_0)$ are valid for {\it any} initial distribution with two levels, whatever the number of patches and their shape. They allow therefore to describe a wider class of situations.

(ii) The variables  $(E,M_0)$ may lead to redundancies because there may exist two (or more) couples    $(E,M_0^{(1)})$ and $(E,M_0^{(2)})$ that correspond to the {\it same}  $(E,f_0)$ and, consequently, to the {\it same} Lynden-Bell state (recall that the Lynden-Bell prediction only depends on $E$ and $f_0$) \footnote{It can be noted that, in case of {\it incomplete} relaxation, the initial conditions $(E,M_0^{(1)})$ and $(E,M_0^{(2)})$ can lead to different results, although the Lynden-Bell theory leads to the same prediction. This could be interesting to check.}. This has been illustrated in \cite{proc,staniscia}.

(iii) More importantly, the use of $M_0$ as an external parameter (instead of $f_0$) leads to physical inconsistencies in the thermodynamical analysis. Indeed, if we work in terms of the variables $(E,M_0)$, the initial value of the distribution $f_0$ becomes a function $f_0=\phi(E,M_0)$ of these variables. As a result, the Lynden-Bell entropy functional
\begin{eqnarray}
\label{lbt9}
S=-\int \biggl\lbrack \frac{\overline{f}}{\phi(E,M_0)}\ln \frac{\overline{f}}{\phi(E,M_0)}\nonumber\\
+\left (1-\frac{\overline{f}}{\phi(E,M_0)}\right )\ln \left (1-\frac{\overline{f}}{\phi(E,M_0)}\right )\biggr\rbrack\, d\theta dv,
\end{eqnarray}
depends not only of the external parameter $M_0$ but also on the energy $E$. This is clearly a very  unconventional situation. Indeed, if we want to apply the standard results recalled above, the entropic functional  can depend on an external parameter but it cannot explicitly depend on the energy. Therefore, these general results \cite{ellis} are not valid for functionals of the form (\ref{lbt9}). In particular, the ``improper'' caloric curve $\beta(E)$ at fixed $M_0$ can display a region of negative specific heat while the proper caloric curve $\beta(E)$ at fixed $f_0$ does not. This is exemplified in Fig. 1(b) of \cite{antovlasov} where the entropy versus energy is plotted at fixed $M_0$. This curve has a convex dip (revealing a negative specific heat region), while the curve $S(E)$ at fixed $f_0$ has no convex dip and the ensembles are equivalent \cite{staniscia2}.

Finally, in the other contexts where the Lynden-Bell theory has been applied \cite{csmnras,staquet,brands,chen,wt,bsj}, the control parameters that have been taken are $E$ and $f_0$. It is therefore important to describe the phase transitions in terms of these parameters as initiated in \cite{epjb}.

\section{Description of caloric curves and phase transitions}
\label{sec_description}

\subsection{Phase diagrams}
\label{sec_pd}

\begin{figure}
\begin{center}
\includegraphics[clip,scale=0.3]{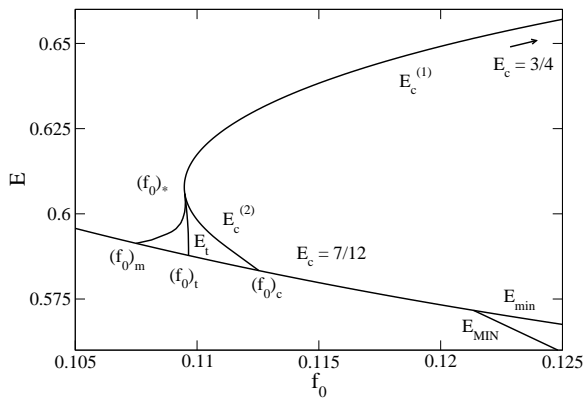}
\caption{Phase diagram in the microcanonical ensemble. We have indicated several representative points. The minimum accessible energy of the homogeneous phase, corresponding to a waterbag (or Fermi) distribution,  is $E_{min}(f_0)=1/(96\pi^2f_0^2)+1/2$ \cite{epjb,proc}. We have also indicated the minimum accessible energy (for homogeneous and inhomogeneous phases) in the case where the initial condition is a rectangular waterbag initial condition. It is equal to $E_{min}(f_0)$ when $f_0<0.12135...$ and to $E_{MIN}(f_0)$ when $f_0>0.12135...$ (see \cite{staniscia} for details). The curves $E_t(f_0)$ and $E_m(f_0)$ have been continued ``by hand'' (due to numerical problems) and may not be correct for small energies (see Appendix \ref{sec_ground}).}
\label{f_0-U_2}
\end{center}
\end{figure}

\begin{figure}
\begin{center}
\includegraphics[clip,scale=0.3]{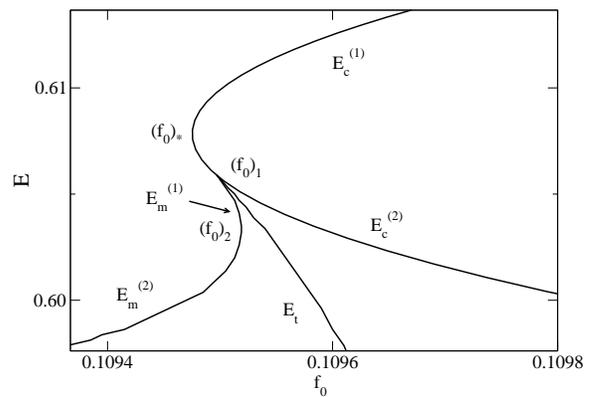}
\caption{Enlargement of the phase diagram in the microcanonical ensemble.}
\label{f_0-U-ZOOM}
\end{center}
\end{figure}

\begin{figure}
\begin{center}
\includegraphics[clip,scale=0.3]{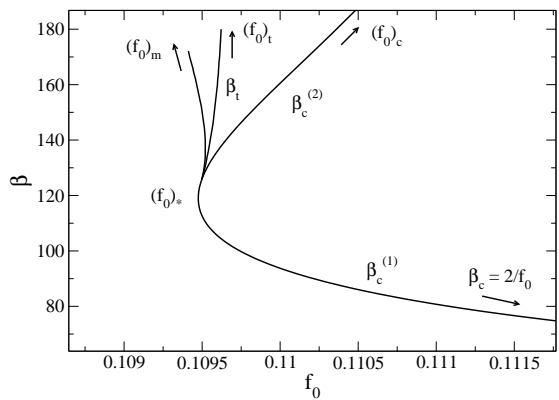}
\caption{Phase diagram in the canonical ensemble.}
\label{f_0-beta}
\end{center}
\end{figure}

\begin{figure}
\begin{center}
\includegraphics[clip,scale=0.3]{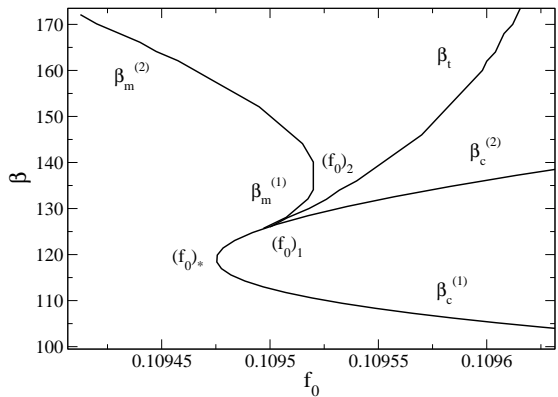}
\caption{Enlargement of the phase diagram in the canonical ensemble.}
\label{f_0-beta_2}
\end{center}
\end{figure}

In Fig. \ref{f_0-U_2}, we reproduce the microcanonical phase diagram obtained in \cite{staniscia}. In Fig. \ref{f_0-U-ZOOM}, we enlarge this diagram close to the turning point of energy $((f_0)_*,E_*)\simeq (0.10947,0.608)$ in order to show that its structure is more complicated than previously thought. Similarly, in Figs. \ref{f_0-beta} and  \ref{f_0-beta_2}, we plot the canonical phase diagram and its enlargement close to the turning point of temperature $((f_0)_*,\beta_*)\simeq (0.10947,118)$. These phase diagrams show that we must consider different regions where the nature of phase transitions changes.

In the following sections, we plot the series of equilibria $\beta(E)$ in seven characteristic  regions of the phase diagram and describe the corresponding phase transitions.  The branches (S) correspond to fully stable states, the branches (M) correspond to metastable states and  the branches (U) correspond  to unstable states. At the end of each subsection, we summarize the nature of phase transitions in the corresponding region by considering only fully stable states.

We will find that the microcanonical and canonical phase transitions are very similar. In fact,  the ensembles differ only in a very small range of parameters. Therefore, we will essentially focus on the microcanonical ensemble and only mention the canonical ensemble in case of ensembles inequivalence.

\subsection{Region 5}
\label{sec_r5}

\begin{figure}
\begin{center}
\includegraphics[clip,scale=0.3]{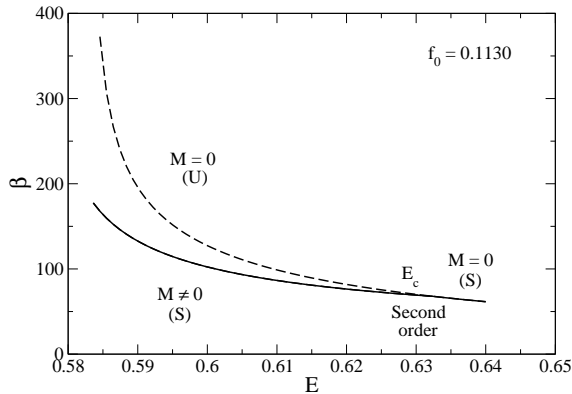}
\caption{Series of equilibria in Region 5 displaying microcanonical and canonical second order phase transitions at $E_c$ and $\beta_c$ respectively. }
\label{beta_u_0.1130}
\end{center}
\end{figure}

\begin{figure}
\begin{center}
\includegraphics[clip,scale=0.3]{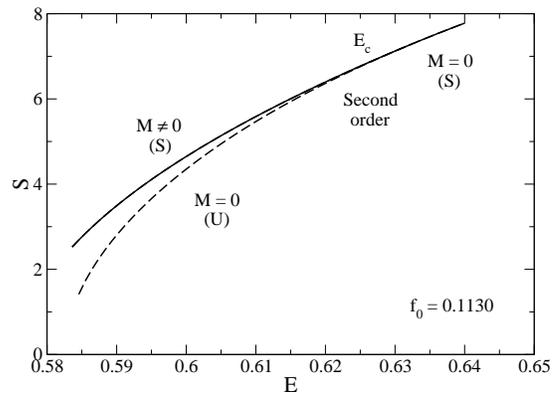}
\caption{Entropy versus energy in Region 5.}
\label{ento_u_0.1130}
\end{center}
\end{figure}

\begin{figure}
\begin{center}
\includegraphics[clip,scale=0.3]{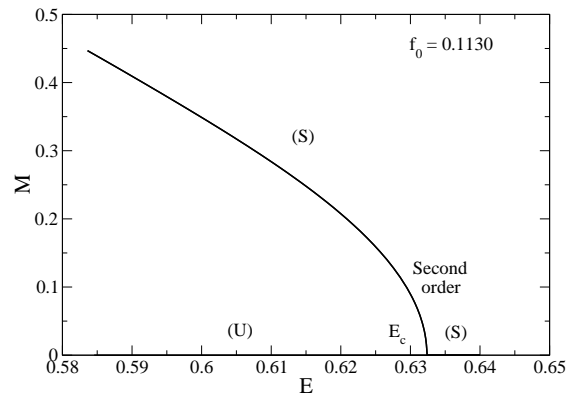}
\caption{Magnetization versus energy in Region 5.}
\label{m_u_0.1130}
\end{center}
\end{figure}

In Figs. \ref{beta_u_0.1130}-\ref{m_u_0.1130} we plot the series of
equilibria in Region 5 corresponding to $f_0>(f_0)_c$ where
$(f_0)_c=1/(2\pi\sqrt{2})\simeq 0.11253954$ (see Fig. \ref{f_0-U_2})
\cite{epjb,staniscia}. Specifically, we consider $f_0=0.1130$.

The homogeneous phase exists at any accessible energy. It is fully stable for $E>E_c$ and unstable for $E<E_c$. The inhomogeneous phase exists for $E<E_c$. It has a higher entropy (see Fig. \ref{ento_u_0.1130}) than the homogeneous phase and it is fully stable. Therefore, the microcanonical caloric curve displays a second order phase transition between homogeneous and inhomogeneous states marked by the discontinuity of $\beta'(E)$ at $E=E_c$. In the entropic curve of Fig. \ref{ento_u_0.1130}, this corresponds to a discontinuity of the second derivative $S''(E)=\beta'(E)$ at $E=E_c$. The magnetization passes from $M=0$ for $E>E_c$ to $M\neq 0$ for $E<E_c$ but remains continuous at the transition (see Fig. \ref{m_u_0.1130}). The discussion is similar in the canonical ensemble.

Region 5: (i) in the MCE, there exists a second order phase transition at $E_c$. (ii) In the CE, there exists a second order phase transition at $\beta_c$. The ensembles are equivalent.

\subsection{Region 4}
\label{sec_r4}

\begin{figure}
\begin{center}
\includegraphics[clip,scale=0.3]{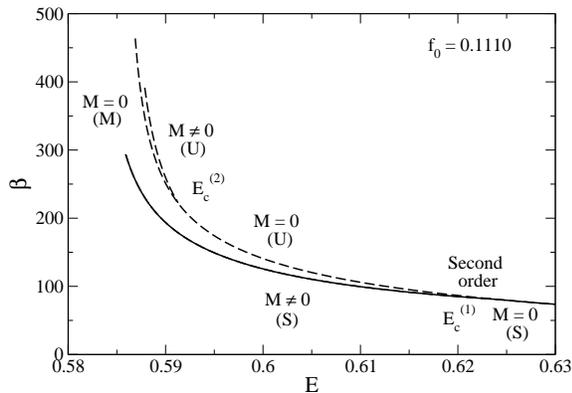}
\caption{Series of equilibria in Region 4 displaying microcanonical and canonical second order phase transitions at  $E_c^{(1)}$ and $\beta_c^{(1)}$ respectively. At $E_c^{(2)}$ and  $\beta_c^{(2)}$, a second inhomogeneous phase appears (but is unstable) while the homogeneous phase becomes metastable.}
\label{beta_u_0.1110}
\end{center}
\end{figure}

In Figs. \ref{beta_u_0.1110}-\ref{m_u_0.1110} we plot the series of equilibria in Region 4 corresponding to $(f_0)_t<f_0<(f_0)_c$ where $(f_0)_t\simeq 0.10965$ and $(f_0)_c\simeq 0.11253954$ (see Fig. \ref{f_0-U_2}). Specifically, we consider $f_0=0.1110$.

The homogeneous phase exists at any accessible energy. It is fully stable for $E>E_c^{(1)}$, unstable for $E_c^{(2)}<E<E_c^{(1)}$ and metastable for $E<E_c^{(2)}$.  A first  inhomogeneous phase exists for $E<E_c^{(1)}$. It has a higher entropy than the homogeneous phase and it is fully stable (see Fig. \ref{entro_u_0.1110}). Therefore, the microcanonical caloric curve displays a second order phase transition between homogeneous and inhomogeneous states marked by the discontinuity of $S''(E)=\beta'(E)$ at $E=E_c^{(1)}$.  The magnetization passes from $M=0$ for $E>E_c^{(1)}$ to $M\neq 0$ for $E<E_c^{(1)}$ but remains continuous at the transition (see Fig. \ref{m_u_0.1110}). A second inhomogeneous phase exists for $E<E_c^{(2)}$. It appears precisely at the energy $E_c^{(2)}$  at which the homogeneous phase becomes metastable. It has a lower entropy $S$ than the homogeneous phase and the first inhomogeneous phase (see Fig. \ref{entro_u_0.1110}) and it is unstable. This branch is clearly visible on the magnetization curve (see Fig. \ref{m_u_0.1110}). The discussion is similar in the canonical ensemble.

Region 4: (i) in the MCE, there exists a second order phase transition at $E_c^{(1)}$. (ii) In the CE, there exists a second order phase transition at $\beta_c^{(1)}$. The ensembles are equivalent.

\begin{figure}
\begin{center}
\includegraphics[clip,scale=0.3]{entro_u_0.1110.eps}
\caption{Entropy versus energy in Region 4.}
\label{entro_u_0.1110}
\end{center}
\end{figure}

\begin{figure}
\begin{center}
\includegraphics[clip,scale=0.3]{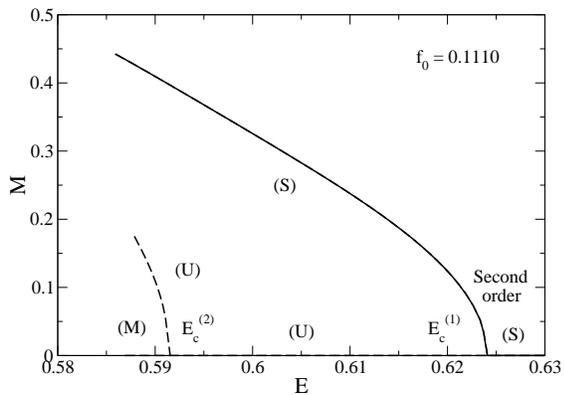}
\caption{Magnetization versus energy in Region 4. The unstable inhomogeneous phase is clearly visible.}
\label{m_u_0.1110}
\end{center}
\end{figure}

\subsection{Region 3-c}
\label{sec_r3c}

\begin{figure}
\begin{center}
\includegraphics[clip,scale=0.3]{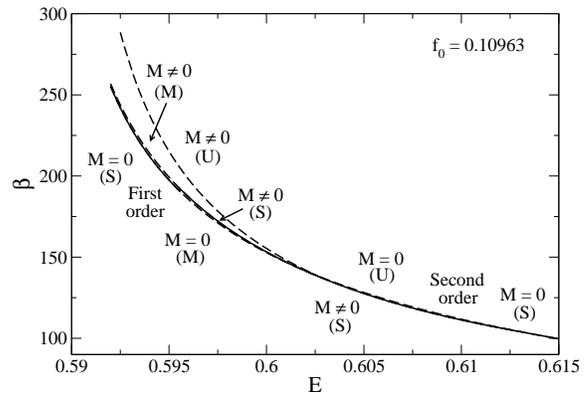}
\caption{Series of equilibria in Region 3-c. It  displays microcanonical and canonical second order phase transitions at $E_c^{(1)}$ and $\beta_c^{(1)}$ respectively. It also displays microcanonical and canonical first order phase transitions at $E_t$ and $\beta_t$ respectively. }
\label{beta_u_0.109630}
\end{center}
\end{figure}

\begin{figure}
\begin{center}
\includegraphics[clip,scale=0.3]{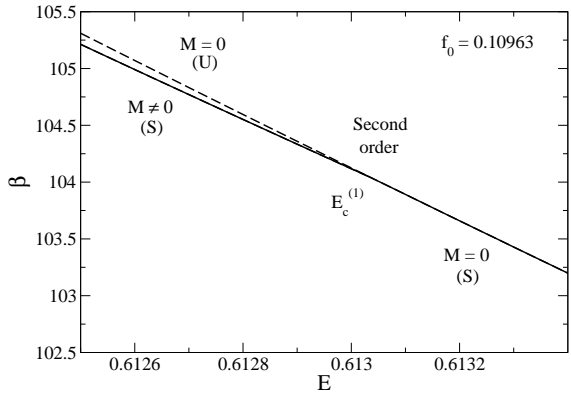}
\caption{Enlargement of Fig. \ref{beta_u_0.109630} in the region of second order phase transition.}
\label{beta_u_0.109630ZOOM2}
\end{center}
\end{figure}

\begin{figure}
\begin{center}
\includegraphics[clip,scale=0.3]{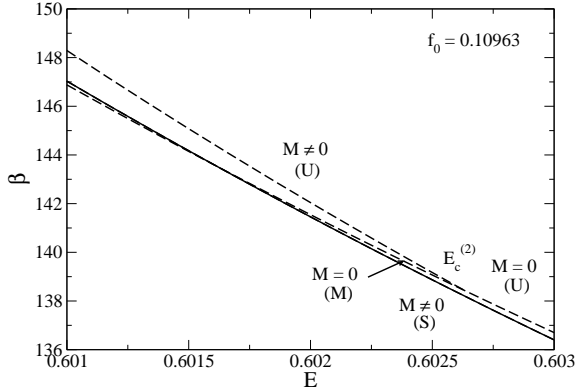}
\caption{Enlargement of Fig. \ref{beta_u_0.109630} near the energy $E_c^{(2)}$ where a second inhomogeneous phase (unstable) appears. At that point, the homogeneous phase becomes metastable. We also note, in passing,  that the temperatures of the  metastable homogeneous phase and of  the fully stable inhomogeneous phase cross each other at some point but this does not signal a change of stability. In particular, the homogeneous phase remains metastable until the energy $E_t$ of first order phase transition (see Fig. \ref{beta_u_0.109630ZOOM1}).}
\label{beta_u_0.109630ZOOMNEW}
\end{center}
\end{figure}

\begin{figure}
\begin{center}
\includegraphics[clip,scale=0.3]{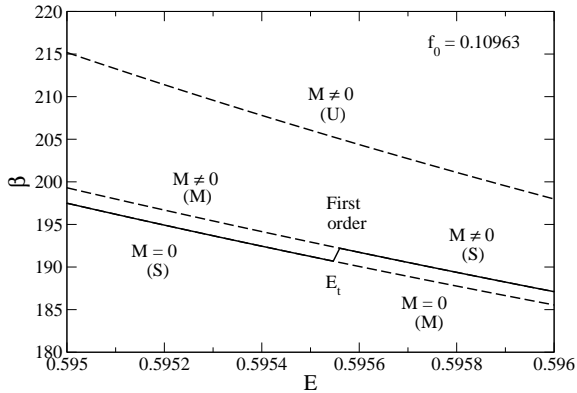}
\caption{Enlargement of Fig. \ref{beta_u_0.109630} in the region of first order phase transition.}
\label{beta_u_0.109630ZOOM1}
\end{center}
\end{figure}

\begin{figure}
\begin{center}
\includegraphics[clip,scale=0.3]{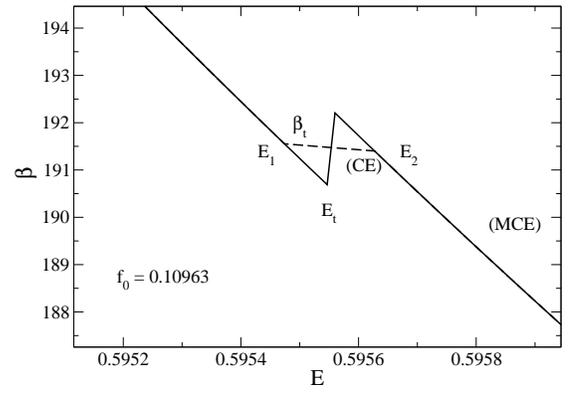}
\caption{Enlargement of Fig. \ref{beta_u_0.109630} in the region of ensembles inequivalence. The full line corresponds to the strict caloric curve in the MCE and the dashed line to the strict caloric curve in the CE. The states between $E_1\simeq 0.59547$ and $E_2\simeq 0.595629$ are stable in the microcanonical ensemble while they are unstable (i.e. inaccessible) in the canonical ensemble. Note that the domain of ensembles inequivalence is almost imperceptible.}
\label{beta_u_0.109630_confr}
\end{center}
\end{figure}

\begin{figure}
\begin{center}
\includegraphics[clip,scale=0.3]{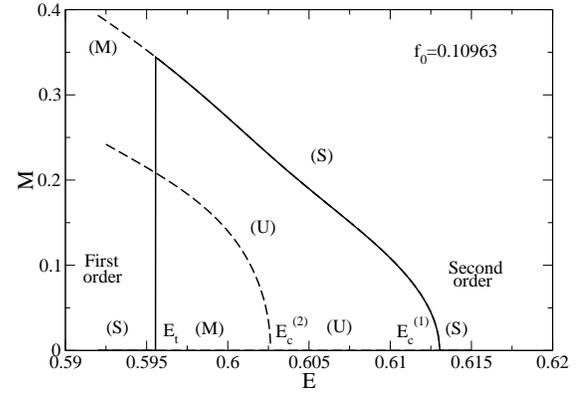}
\caption{Magnetization versus energy in Region 3-c showing clearly the first and second order phase transitions as well as the second (unstable) inhomogeneous branch. }
\label{m_u_0.109630}
\end{center}
\end{figure}

In Figs. \ref{beta_u_0.109630}-\ref{m_u_0.109630}, we plot the series of equilibria in Region 3-c corresponding to $(f_0)_2<f_0<(f_0)_t$ where $(f_0)_2\simeq 0.109519$ and $(f_0)_t\simeq 0.10965$ (see Fig. \ref{f_0-U-ZOOM}). Specifically, we consider $f_0=0.10963$.

The homogeneous phase exists at any accessible energy. It is fully stable for $E>E_c^{(1)}$, unstable for $E_c^{(2)}<E<E_c^{(1)}$, metastable for $E_{t}<E<E_c^{(2)}$ and fully stable for $E<E_t$.  A first  inhomogeneous phase exists for $E<E_c^{(1)}$. It is fully stable for $E_{t}<E<E_c^{(1)}$ and metastable for $E<E_t$. Indeed, it has a higher entropy than the homogeneous phase for $E_{t}<E<E_c^{(1)}$ and a lower entropy for $E<E_t$. Therefore, the microcanonical caloric curve displays a second order phase transition between homogeneous and inhomogeneous states marked by the discontinuity of $\beta'(E)=S''(E)$ at $E=E_c^{(1)}$ (see Fig. \ref{beta_u_0.109630ZOOM2}) and a first order phase transition between homogeneous and inhomogeneous states marked by the discontinuity of $\beta(E)=S'(E)$ at $E=E_t$ (see Fig. \ref{beta_u_0.109630ZOOM1}). The magnetization of the fully stable branch passes from $M=0$ to $M\neq 0$ at $E=E_c^{(1)}$ but remains continuous, and it passes from $M\neq 0$ to $M=0$ at $E=E_t$ by being discontinuous (see Fig. \ref{m_u_0.109630}). We note that the first order phase transition is hardly visible on the caloric curve $\beta(E)$ whereas it is clearly visible on the magnetization curve $M(E)$. A second inhomogeneous phase exists for $E<E_c^{(2)}$. It appears precisely at the energy $E_c^{(2)}$  at which the homogeneous phase becomes metastable. It has a lower entropy $S$ than the homogeneous phase and the first inhomogeneous phase and it is unstable. This branch is clearly visible on the magnetization curve of Fig. \ref{m_u_0.109630}. The discussion  is similar in the canonical ensemble.

Region 3-c: (i) In MCE, there exists a second order phase transition at $E_c^{(1)}$ and a first order phase transition at $E_t$ \footnote{A similar situation has been observed in \cite{cc} for polytropic distributions.}. (ii) In CE, there exists a second order phase transition at $\beta_c^{(1)}$ and a first order phase transition at $\beta_t$. For $0.595477\le E\le 0.595629$, the  ensembles are inequivalent (see Fig. \ref{beta_u_0.109630_confr}). However, this concerns a strikingly narrow range of energies.

\subsection{Region 3-b}
\label{sec_r3b}

\begin{figure}
\begin{center}
\includegraphics[clip,scale=0.3]{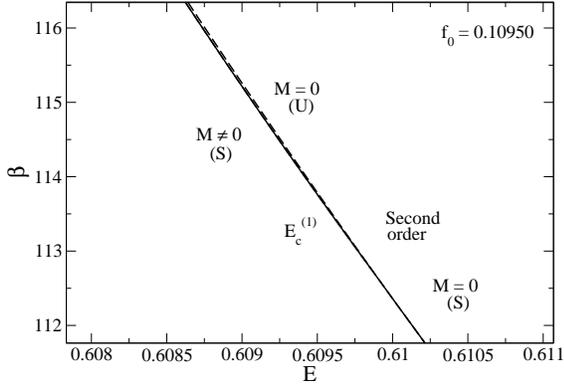}
\caption{Series of equilibria in Region 3-b near the point of second order phase transition.}
\label{beta_u_0.10950ZOOM2}
\end{center}
\end{figure}

\begin{figure}
\begin{center}
\includegraphics[clip,scale=0.3]{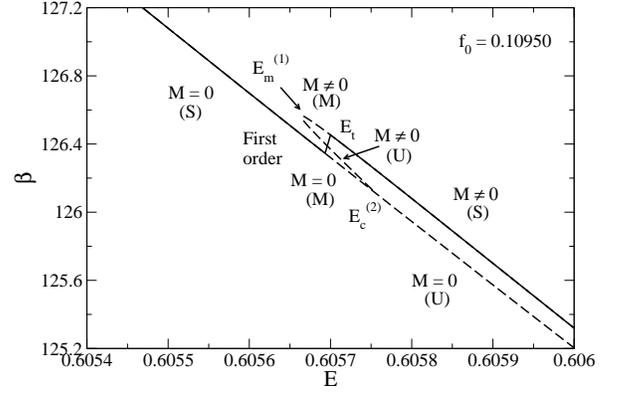}
\caption{Series of equilibria in Region 3-b near the point of first order phase transition. We also see the disappearance of the inhomogeneous phases for $E<E_m^{(1)}$. In particular, the energy $E_m^{(1)}$ can be interpreted as a spinodal point at which the inhomogeneous metastable branch disappears.}
\label{beta_u_0.10950ZOOM1}
\end{center}
\end{figure}

\begin{figure}
\begin{center}
\includegraphics[clip,scale=0.3]{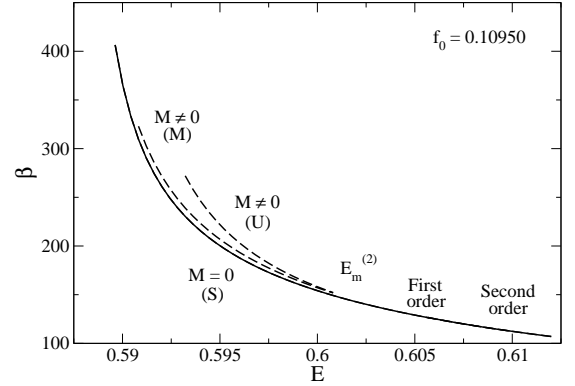}
\caption{Series of equilibria in Region 3-b. We see the re-appearance of the inhomogeneous phases for $E<E_m^{(2)}$. The energy $E_m^{(2)}$ can also be interpreted as a spinodal point.}
\label{beta_u_0.10950ZOOMNEW}
\end{center}
\end{figure}

\begin{figure}
\begin{center}
\includegraphics[clip,scale=0.3]{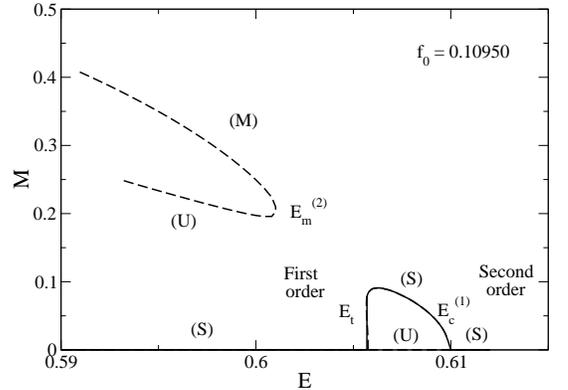}
\caption{Magnetization versus energy in Region 3-b.}
\label{m_u_0.10950}
\end{center}
\end{figure}

\begin{figure}
\begin{center}
\includegraphics[clip,scale=0.3]{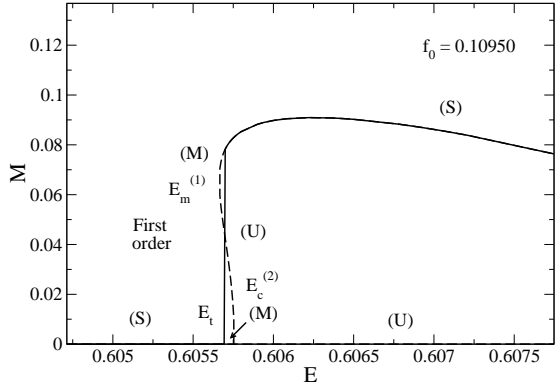}
\caption{Enlargement of Fig. \ref{m_u_0.10950} in the region of first order phase transition.}
\label{m_u_0.10950ZOOM}
\end{center}
\end{figure}

In Figs. \ref{beta_u_0.10950ZOOM2}-\ref{m_u_0.10950ZOOM}, we plot the series of equilibria in Region 3-b corresponding to $(f_0)_1<f_0<(f_0)_2$ where $(f_0)_1\simeq 0.109497$ and  $(f_0)_2\simeq 0.109519$ (see Fig. \ref{f_0-U-ZOOM}). Specifically, we consider $f_0=0.10950$.

The homogeneous phase exists at any
accessible energy. It is fully stable for $E>E_c^{(1)}$,
unstable for $E_c^{(2)}<E<E_c^{(1)}$, metastable for
$E_{t}<E<E_c^{(2)}$ and fully stable for $E<E_t$.  A first
inhomogeneous phase exists for $E_{m}^{(1)}<E<E_c^{(1)}$ and $E<E_m^{(2)}$ (it does not exist between $E_m^{(2)}$ and $E_{m}^{(1)}$). It is fully stable for $E_{t}<E<E_c^{(1)}$ and metastable for $E_{m}^{(1)}<E<E_t$ and for $E<E_m^{(2)}$.
Therefore, the microcanonical caloric curve displays a second order phase transition between homogeneous and inhomogeneous states marked by the discontinuity of $\beta'(E)=S''(E)$ at $E=E_c^{(1)}$ and a first order phase transition between homogeneous and inhomogeneous states marked by the discontinuity of $\beta(E)=S'(E)$ at $E=E_t$. A second inhomogeneous phase exists for $E_{m}^{(1)}<E<E_c^{(2)}$ and for  $E<E_m^{(2)}$ (it does not exist between $E_m^{(2)}$ and $E_{m}^{(1)}$). It appears precisely at the energy $E_c^{(2)}$ at which the homogeneous phase becomes metastable. This second inhomogeneous phase is always unstable. The first and second order phase transitions are better visible on the magnetization curves of Figs. \ref{m_u_0.10950} and \ref{m_u_0.10950ZOOM}. The discussion  is similar in the canonical ensemble.

Region 3-b: (i) In MCE, there exists a second order phase transition at $E_c^{(1)}$ and a first order phase transition at $E_t$. (ii) In CE, there exists a second order phase transition at $\beta_c^{(1)}$ and a first order phase transition at $\beta_t$. As in region 3-c, there exists a tiny region of ensembles inequivalence.

\subsection{Region 3-a}
\label{sec_r3a}

\begin{figure}
\begin{center}
\includegraphics[clip,scale=0.3]{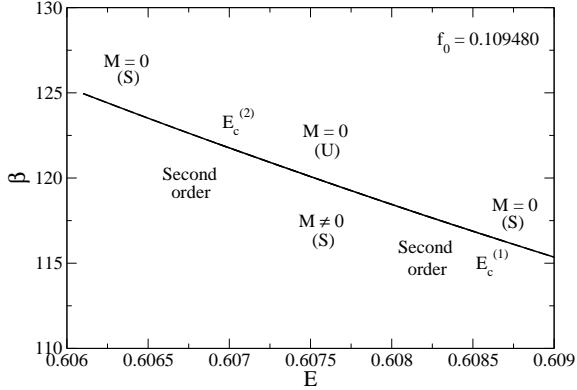}
\caption{Series of equilibria in Region 3-a. It  displays two microcanonical and canonical second order phase transitions at  $E_c^{(1)}$, $E_c^{(2)}$  and at $\beta_c^{(1)}$, $\beta_c^{(2)}$ respectively.}
\label{beta_u_0.109480}
\end{center}
\end{figure}

\begin{figure}
\begin{center}
\includegraphics[clip,scale=0.3]{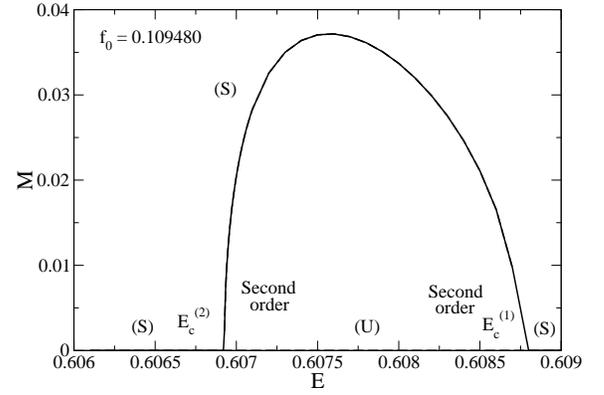}
\caption{Magnetization versus energy in Region 3-a. It  displays two microcanonical second order phase transitions at  $E_c^{(1)}$ and  $E_c^{(2)}$.}
\label{m_u_0.109480}
\end{center}
\end{figure}

\begin{figure}
\begin{center}
\includegraphics[clip,scale=0.3]{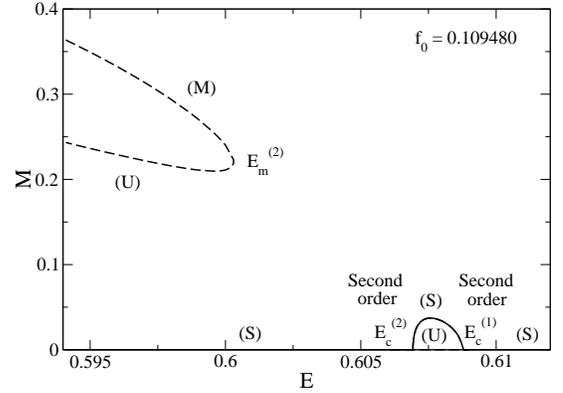}
\caption{Extension of Fig. \ref{m_u_0.109480} showing the reappearance of the inhomogeneous phase for $E<E_m^{(2)}$.}
\label{m_u_0.109480ENLARGED}
\end{center}
\end{figure}

In Figs. \ref{beta_u_0.109480}-\ref{m_u_0.109480ENLARGED}, we plot the series of equilibria in Region 3-a corresponding to $(f_0)_*<f_0<(f_0)_1$ where   $(f_0)_*\simeq 0.10947$ and $(f_0)_1\simeq 0.109497$ (see Fig. \ref{f_0-U-ZOOM}). Specifically, we consider $f_0=0.109480$.

The homogeneous phase exists at any accessible energy. It is fully stable for $E>E_c^{(1)}$, unstable for $E_c^{(2)}<E<E_c^{(1)}$, and fully stable for $E<E_c^{(2)}$. A first inhomogeneous phase exists for $E_c^{(2)}<E<E_c^{(1)}$ and for $E<E_m^{(2)}$  (it does not exist for $E_m^{(2)}<E<E_c^{(2)}$). It is fully stable for $E_c^{(2)}<E<E_c^{(1)}$ and metastable for $E<E_m^{(2)}$. Therefore, the microcanonical caloric curve displays two second order phase transitions between homogeneous and inhomogeneous states marked by the discontinuity of $\beta'(E)=S''(E)$ at $E=E_c^{(1)}$ and $E=E_c^{(2)}$.   The magnetization of the fully stable branch passes from $M=0$ to $M\neq 0$ at $E=E_c^{(1)}$ and from $M\neq 0$ to $M=0$ at $E=E_c^{(2)}$, but remains continuous at the transition (see Fig. \ref{m_u_0.109480}). We note that the second order phase transitions are hardly visible on the caloric curve $\beta(E)$ whereas they are clearly visible on the magnetization curve $M(E)$. A second inhomogeneous phase exists for $E<E_m^{(2)}$ and it is unstable.  The discussion  is similar in the canonical ensemble.

Region 3-a: (i) In MCE, there exists two second order phase transitions at $E_c^{(1)}$ and $E_c^{(2)}$; (ii) in CE, there exists two second order phase transitions at $\beta_c^{(1)}$ and $\beta_c^{(2)}$. The ensembles are equivalent.

\subsection{Region 2}
\label{sec_r2}

\begin{figure}
\begin{center}
\includegraphics[clip,scale=0.3]{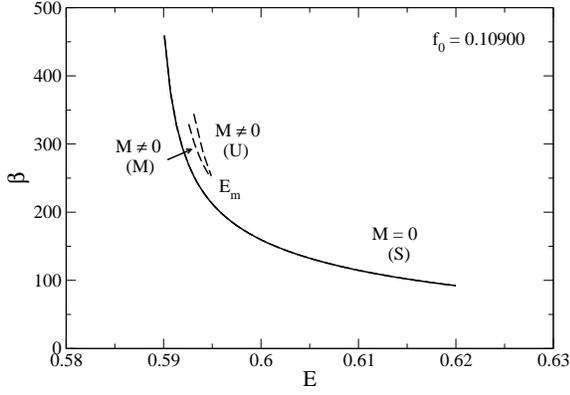}
\caption{Series of equilibria in Region 2. There is no phase transition but the sudden appearance of a metastable inhomogeneous branch, accompanied by an unstable inhomogeneous branch, at $E=E_m$.}
\label{beta_u_0.10900}
\end{center}
\end{figure}

\begin{figure}
\begin{center}
\includegraphics[clip,scale=0.3]{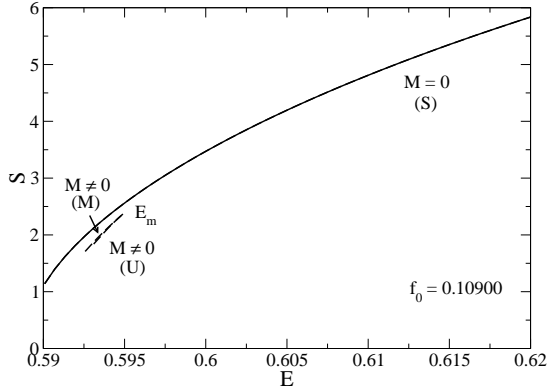}
\caption{Entropy versus energy  in Region 2.}
\label{entro_u_0.10900}
\end{center}
\end{figure}

\begin{figure}
\begin{center}
\includegraphics[clip,scale=0.3]{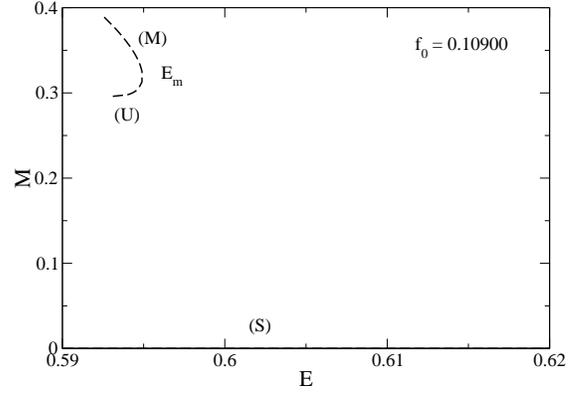}
\caption{Magnetization versus energy in Region 2.}
\label{m_u_0.10900}
\end{center}
\end{figure}

In Figs. \ref{beta_u_0.10900}-\ref{m_u_0.10900}, we plot the series of equilibria in Region 2 corresponding to $(f_0)_m<f_0<(f_0)_*$ where  $(f_0)_m\simeq 0.1075$ and  $(f_0)_*\simeq 0.10947$  (see  Fig. \ref{f_0-U_2}). Specifically, we consider $f_0=0.10900$.

The homogeneous phase exists at any accessible energy and it is fully stable.  Therefore, the microcanonical caloric curve does not display any phase transition and is made of homogeneous states. Two  inhomogeneous phases appear for $E<E_{m}$, one being metastable and the other unstable. The metastable phase has a lower entropy than the homogeneous phase and the unstable phase has a lower entropy than the metastable phase (see Fig. \ref{entro_u_0.10900}). These different phases can also be seen on the magnetization (order parameter) curve of Fig. \ref{m_u_0.10900}. The discussion  is similar in the canonical ensemble.

Region 2: There is no phase transition and the ensembles are equivalent.

\subsection{Region 1}
\label{sec_r1}

\begin{figure}
\begin{center}
\includegraphics[clip,scale=0.3]{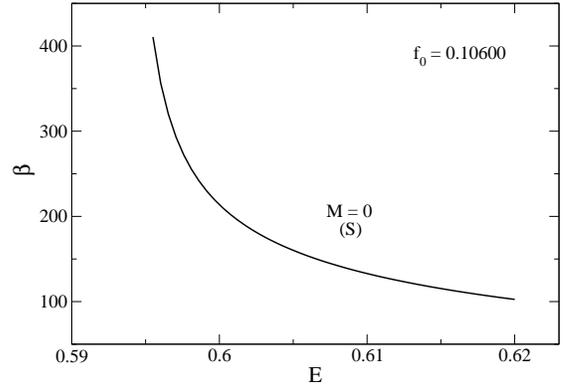}
\caption{Series of equilibria in Region 1.}
\label{beta_u_0.10600}
\end{center}
\end{figure}

\begin{figure}
\begin{center}
\includegraphics[clip,scale=0.3]{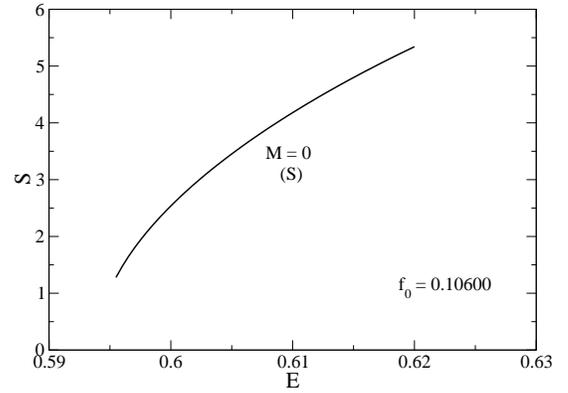}
\caption{Entropy versus energy curve in Region $1$.}
\label{entro_u_0.10600}
\end{center}
\end{figure}

In Figs. \ref{beta_u_0.10600} and \ref{entro_u_0.10600}, we plot the series of equilibria in Region 1 corresponding to $f_0<(f_0)_m$ where  $(f_0)_m\simeq 0.1075$ (see  Fig. \ref{f_0-U_2}). Specifically, we consider $f_0=0.10600$.

The homogeneous phase exists at any accessible energy and it is fully stable. There is no inhomogeneous phase.  Therefore, the microcanonical caloric curve does not display any phase transition and is made of homogeneous states (see Figs. \ref{beta_u_0.10600} and \ref{entro_u_0.10600}). The discussion  is similar in the canonical ensemble.

Region 1: There is no phase transition and the ensembles are equivalent.

\section{Discussion}
\label{sec_discussion}

Let us summarize the different results obtained in the previous analysis:

(i) Decreasing $f_0$, the system successively  exhibits one second order phase transition (Regions 5 and 4),  one second order and one first order phase transition (Regions 3-c and 3-b), two second order phase transitions (Region 3-a), and no phase transition (Regions 2 and 1).

(ii) There exists a {\it  tricritical point} corresponding to the passage from a first order phase transition to a second order phase transition. It is located at $((f_0)_1, E_1, \beta_1)\simeq (0.109497, 0.6059, 125)$.

(iii) The sudden appearance of two second order phase transitions at the turning  point $((f_0)_*,E_*,\beta_*)\simeq (0.10947, 0.608, 118)$ is sometimes called {\it second order azeotropy} \cite{bb}.

(iv) For $(f_0)_*<f_0<(f_0)_c$, there is a phenomenon of {\it phase reentrance} concerning the homogeneous phase \cite{epjb,staniscia}. As we reduce the energy, the homogeneous phase is successively stable, unstable and stable (or metastable) again. This phenomenon  is basically due to the turning point of the energy curve $E_c(f_0)$ at $f_0=(f_0)_*$. It is therefore associated with the second order azeotropy.

(v) For $(f_0)_*<f_0<(f_0)_2$, there is a phenomenon of phase reentrance concerning the inhomogeneous phase.  As we reduce the energy, the inhomogeneous phase is stable (or metastable), then it disappears, and it finally reappears as a metastable state. This phenomenon is basically due to the turning point of the energy curve $E_m(f_0)$ at $f_0=(f_0)_2$. It is located at  $((f_0)_2, E_2, \beta_2)\simeq (0.109519,0.603,137)$.

(vi) The tricritical point $((f_0)_1,E_1)$ separating first and second order phase transitions is located between the turning points of the $E_c(f_0)$ and $E_m(f_0)$ curves.

(vii) In Regions 3-c and 3-b, there is a very small zone of ensembles inequivalence associated with the  first order phase transitions.

In conclusion, the out-of-equilibrium phase transitions of the HMF model predicted by the Lynden-Bell theory lead to a rich and interesting phase diagram. It is striking that everything happens in a very narrow range of parameters $(f_0)_m\simeq 0.1075<f_0<(f_0)_c\simeq 0.11253954$, although $f_0$ can take in principle all positive values. A similar observation has been made previously in other studies of phase transitions in systems with long-range interactions \cite{beg,ring,cd,cc}. The branches corresponding to the different phases are very close to each other in the series of equilibria $\beta(E)$ and in the entropic curves $S(E)$. This shows in particular that all the phases have almost the same entropy, even the unstable ones. However, the branches appear to be well separated in the kinetic caloric curve  $\beta_{kin}(E)$ \cite{staniscia2} and in the magnetization curve $M(E)$.

\section{Numerical simulations}
\label{sec_numerics}

\begin{figure}
\begin{center}
\includegraphics[clip,scale=0.3]{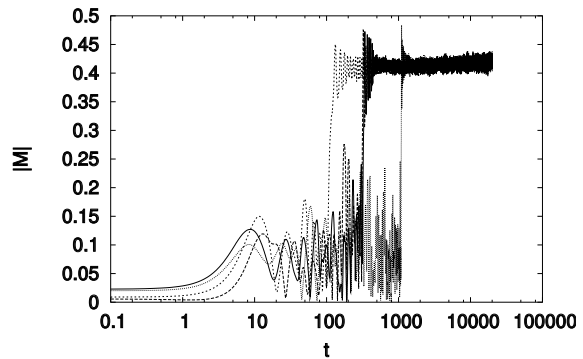}
\caption{Magnetization as a function of time for a system of $N = 20000$ particles at energy $E = 0.5901$. The four curves are different realizations of
 the same initial distribution with $f_0 = 0.1097$. The initial condition is a rectangular waterbag distribution with magnetization $M_0\simeq 0.01413$ (see Eq. (28) of \cite{staniscia}).}
\label{meta1}
\end{center}
\end{figure}

The numerical results obtained in \cite{staniscia} show that the phase
diagram deduced from the Lynden-Bell theory predicts the right
phenomenology, even in the small and quite complex region located
around the tricritical point. In this section, we show that, in the
region of metastability (according to Lynden-Bell's theory), the
system displays the usual dynamical behavior of systems in a
metastable state: a ``lethargic'' evolution during which the system is
trapped in a given macrostate (metastable), followed by a sudden jump
in a different macrostate (fully stable). To the best of our
knowledge, this is the first evidence of metastability during a QSS.
Our results are obtained from a set of molecular dynamics simulations
carried out after having prepared the system in a state belonging to
the region of the ($f_0$,$E$) plane close to the first order
transition line. For the chosen parameters $(f_0,E)=(0.1097,0.5901)$,
the homogeneous phase is metastable and the inhomogeneous phase is
fully stable (see Fig. \ref{f_0-U_2}). Following the temporal
evolution of the magnetization (see Fig. \ref{meta1}), one first
observes the spontaneous relaxation of the system in the unmagnetized
phase (metastable), followed by a sudden jump in the magnetized state
(fully stable).

Different runs with the same initial distribution show that the jump
occurs at random times. This is in agreement with the ordinary
behavior of metastable states, where the time of the jump depends on
the particular ``realization''. The average time at which the jump
occurs depends on the size of the system and increases with $N$.  This
indicates that ``collisions'' (finite $N$ effects) play some role in
the dynamics. This is relatively unexpected since the regime that we
are exploring corresponds to the QSS regime where the Vlasov equation
should be applicable. In all the numerical simulations that we have
run, the system spontaneously relaxes towards the metastable state; it
never directly reaches the fully stable state. The selection of the
QSS, among these two states, obviously depends on a complicated notion
of {\it basin of attraction}. Our initial condition consists in a
rectangular waterbag distribution which has a very small magnetization
$M_0\simeq 0.01744$ since the chosen energy is close to the minimum
energy state $E_{min}(f_0)\simeq 0.58766$ which is a waterbag
distribution with vanishing magnetization
\cite{epjb,staniscia}.  It is likely that this initial condition
belongs to the basin of attraction of the homogeneous metastable
state. It is possible that changing the initial condition (still with
two levels and with the same $f_0$ and $E$ but no more waterbag) so as
to increase the magnetization $M_0$ may help the system to access
directly to the inhomogeneous (fully stable) state. On the other hand,
on the other side of the first order transition line (i.e. for smaller
values of $f_0$), the homogeneous state becomes fully stable while the
inhomogeneous state is metastable. In that case, starting from a
rectangular waterbag initial condition, the system relaxes towards the
homogeneous state (since it belongs to its basin of attraction) and
stays there during the whole QSS regime since it is now fully stable
(numerical simulations not shown). Additional numerical simulations
are necessary to get a more general picture of the QSS metastability
for different values of the control parameters ($f_0,E)$ and different
types of initial conditions.

\section{Conclusion}
\label{sec_conclusion}

In this paper we have explored the phase diagram obtained by applying
Lynden-Bell's statistical theory to the HMF model. We have found a
richer phenomenology in the ($f_0$,$E$) plane \cite{epjb,staniscia}
than in the ($M_0$,$E$) plane \cite{tri,antovlasov}. We have also
explained that the proper external parameter to use in the Lynden-Bell
theory is $f_0$, not $M_0$. The choice of the proper control
parameters has deep consequences on the thermodynamical analysis.

The HMF model is a system in which the Lynden-Bell theory works
relatively well (in contrast to astrophysical systems for which it was
initially devised \cite{lb}). In particular, the phenomenon of phase
reentrance that is predicted on the basis of this theory \cite{epjb}
has been successfully reproduced in \cite{staniscia}. Numerical
evidence of first and second order phase transitions has also been
given in \cite{staniscia} in agreement with theory \footnote{We should
be careful, however, that Staniscia {\it et al.}  \cite{staniscia} do
not determine the detailed distribution functions $f(\theta,v)$ of the
inhomogeneous states. It is possible that the Lynden-Bell theory
predicts the correct phase transitions in terms of the magnetization
although the distribution functions are not exactly given by
Eq. (\ref{lbt6}). This would be interesting to check in future
works.}. This is remarkable because all these interesting features
occur in a very small region of the phase diagram (typically
$(f_0)_m\simeq 0.1075<f_0<(f_0)_c\simeq 0.11253954$). In this sense, the
Lynden-Bell prediction is not only qualitative but in fact extremely
accurate!

There are, however, cases where the Lynden-Bell theory fails. In the
numerical study of \cite{staniscia}, some discrepancies with the
Lynden-Bell prediction were reported. In particular, unmagnetized
states are observed in the {\it a priori} magnetized region leading to
a second reentrant phase. Inversely, magnetized states are observed in
the {\it a priori} unmagnetized region. On the other hand, the
Lynden-Bell theory cannot explain the region of negative kinetic
specific heat observed numerically by Antoni \& Ruffo \cite{ar} and
Latora {\it et al.} \cite{latora}. These authors start from an initial
condition with magnetization $M_0=1$ in which  all the particles are
located at $\theta=0$. The initial distribution function $f_0$ is
infinite corresponding to the dilute (or non degenerate) limit of the
Lynden-Bell theory in which the predicted QSS coincides with the
Boltzmann distribution. Now, the results of numerical simulations
\cite{ar,latora} are inconsistent with the Boltzmann (hence
Lynden-Bell) distribution in the region of negative kinetic specific
heats. This means that violent relaxation is incomplete
\cite{incomplete} and that the system is trapped in a stable steady
state of the Vlasov equation that is not the most mixed
(i.e. Lynden-Bell) state \cite{epjb}. Recently, Chavanis \& Campa
\cite{cc} have investigated the Vlasov dynamical stability of
polytropic (or Tsallis) distributions and argued that polytropes with
an index close to $n=1$ could provide an explanation of the curious
anomalies observed in \cite{ar,latora}. In this work, the polytropic
distributions are justified by a lack of ergodicity and by incomplete
relaxation. It would be interesting to extend their analysis
(restricted so far to initial conditions with magnetization $M_0=1$)
so as to cover a wider range of parameters and see whether it can
explain similarly the anomalies reported in \cite{staniscia}.

Very recently, a mathematical ``tour de force'' has been accomplished
by Mouhot \& Villani \cite{mv} who rigorously proved that systems with
long-range interactions described by the Vlasov equation possess some
asymptotic ``stabilization'' property in large time, although the
Vlasov equation is time-reversible. More precisely, they show that if
a stable steady state of the Vlasov equation is slightly perturbed,
the perturbation converges in a {\it weak sense} towards a steady
distribution through phase mixing without the help of any extra
diffusion or ensemble averaging. This is refered to as {\it nonlinear
Landau damping}. This is a very important work that shades new light
on the process of phase mixing and, consequently, on the nature of
QSSs. However, these authors criticize the Lynden-Bell approach
arguing that there is no ``universal'' large time behavior of the
solutions of the Vlasov equation in terms of just the conservation
laws and the initial datum. In their words: ``This seems to be bad
news for the statistical theory of the Vlasov equation pioneered by
Lynden-Bell''. Although it is clear that the Lynden-Bell theory has
some limitations due to incomplete relaxation (lack of
mixing/ergodicity) \cite{incomplete}, our series of works related to
the HMF model \cite{epjb,staniscia,staniscia2}, including the present
effort, shows that the Lynden-Bell approach is able to make accurate
predictions that are confirmed by direct numerical
simulations. Therefore, the Lynden-Bell theory remains a valuable tool
even if it is difficult to specify its general domain of validity. In
fact, Mouhot \& Villani \cite{mv} do not totally reject this
statistical approach and point out limitations in the application of
their results. In particular, their theory is based on smooth
functions (which is not the norm in statistical theories) and Landau
damping is a thin effect which might be neglected when it comes to
predict the ``final'' state in a ``turbulent'' situation (which is
precisely the aim of Lynden-Bell's statistical theory).  The subject
is certainly not closed and should lead again to interesting findings
and fruitful discussions.

\appendix

\section{The ground state}
\label{sec_ground}

In this Appendix, we briefly discuss the ground state of the
Lynden-Bell distribution (analogous to the Fermi-Dirac
distribution) and its connection with the phase diagram of
Fig. \ref{f_0-U_2}.

\begin{figure}
\begin{center}
\includegraphics[clip,scale=0.3]{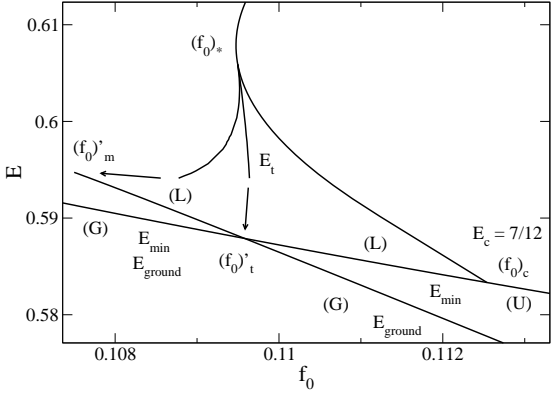}
\caption{Phase diagram close to the ground state. The upper full lines correspond to states that have been actually computed. Arrows give the directions towards which these branches should tend. The lower line denoted $E_{min}$ (like in Fig. \ref{f_0-U_2}) corresponds to the spatially homogeneous waterbag distribution as explained in \cite{epjb,staniscia}. Its energy is $E_{min}=1/(96\pi^2 f_0^2)+1/2$. It is the global energy minimum for $f_0<(f_0)'_t$, a local energy minimum for $(f_0)'_t<f_0<(f_0)_c$ and an unstable saddle point for $f_0>(f_0)_c$. The other lower line  corresponds to the inhomogeneous waterbag distribution as explained in \cite{preparation}. It starts at $f_0=(f_0)'_m$ (corresponding to an energy $E'_m\simeq 0.59473$) and tends to $E=0$ for $f_0\rightarrow +\infty$. It is a local energy minimum for $(f_0)'_m<f_0<(f_0)'_t$ and a global energy minimum for $f_0>(f_0)'_t$. The unstable inhomogeneous waterbag distribution  has not been represented (see \cite{preparation} for details). }
\label{ground}
\end{center}
\end{figure}

\begin{figure}
\begin{center}
\includegraphics[clip,scale=0.3]{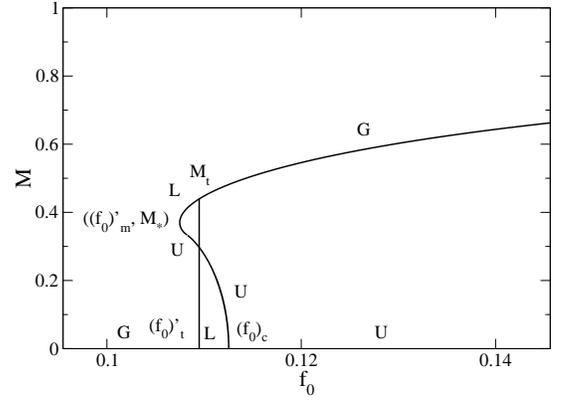}
\caption{Magnetization of the minimum energy state (ground state) as a function of the maximum value of the distribution $f_0$ (taken from \cite{preparation}). It exhibits a first order phase transition at $(f_0)'_t$. The homogeneous waterbag distribution is the global energy minimum for $f_0<(f_0)'_t$, a local energy minimum for $(f_0)'_t<f_0<(f_0)_c$ and an unstable saddle point for $f_0>(f_0)_c$. The inhomogeneous waterbag distribution exists only for $f_0>(f_0)'_m$. It is a local energy minimum for $(f_0)'_m<f_0<(f_0)'_t$ (corresponding to $M_*\simeq 0.37<M<M_t\simeq 0.44$) and the global energy minimum for $f_0>(f_0)'_t$ (corresponding to 
$M>M_t$). It is an unstable saddle point for $(f_0)'_m<f_0<(f_0)_c$ and
$M<M_*$). }
\label{bVSmuN0.5}
\end{center}
\end{figure}

For a given value of $f_0$, the minimum energy state corresponds to a
Fermi distribution at $T=0$, i.e. a (possibly 
spatially inhomogeneous) waterbag distribution. Such a distribution
is equivalent to a polytrope of index $n=1/2$ \cite{cc}. Its
structure and stability are described in detail in
\cite{cc,preparation}. Here, we only give the final
results of the analysis (see Figs. \ref{ground}
and \ref{bVSmuN0.5}) and refer the reader to \cite{cc,preparation} for more details. The global minimum energy state (G) is a
spatially homogeneous waterbag distribution for $f_0<(f_0)'_t\simeq
0.109579$ and a spatially inhomogeneous waterbag distribution for
$f_0>(f_0)'_t$. On the other hand, the spatially inhomogeneous
waterbag distribution is a local minimum energy state (L) for
$(f_0)'_m\simeq 0.1075<f_0<(f_0)'_t$ and the spatially homogeneous
waterbag distribution is a local minimum energy state (L) for
$(f_0)'_t<f_0<(f_0)_c=1/(2\pi\sqrt{2})$. For
$f_0<(f_0)'_m$ (spinodal point), no spatially inhomogeneous waterbag
distribution exists and for $f_0>(f_0)_c$ the spatially homogeneous
waterbag distribution is an unstable saddle point of
energy.

We emphasize that the specific form of the initial condition may
constrain the accessible range of energies. For example, for a
rectangular waterbag initial distribution, the minimum accessible
energy $E_{MIN}(f_0)$ is strictly larger than the ground state
$E_{ground}(f_0)$ for $f_0>(f_0)'_t$ (see
Fig. \ref{f_0-U_2}). Of course, smaller energies can be achieved by
other types of initial conditions.

It is likely that the point $(f_0)'_t\simeq 0.109579$ should coincide
with the transition point $(f_0)_t$ in the phase diagram of
Fig. \ref{f_0-U_2} although we gave a different value $(f_0)_t\simeq
0.10965$ in Sec. \ref{sec_description}. In fact, as we indicated in
the caption of Fig. \ref{f_0-U_2}, the curve $E_t(f_0)$ has been
continued ``by hand'' for small energies so that the value
$(f_0)_t\simeq 0.10965$ is not firmly established and may be
incorrect. The points that have been actually computed are shown in
Fig. \ref{ground}. It is likely that the real curve $E_t(f_0)$ tends
to the point $((f_0)'_t,E_t((f_0)'_t))\simeq (0.109579, 0.587896)$. If
this picture is correct, it implies that the curve $E_t(f_0)$ is
multivalued in some range of parameters $\lbrack (f_0)'_t,
(f_0)_{new}\rbrack$ (say). Indeed, some of the computed points have
values of $f_0$ larger than $(f_0)'_t$ so that the curve must turn
back. This yields an even more complex phase diagram with an
additional phase reentance. Indeed, decreasing the energy in the range
$\lbrack (f_0)'_t, (f_0)_{new}\rbrack$, the homogeneous phase is
successively stable, unstable, metastable, stable, and metastable
again. On the other hand, the inhomogeneous phase is inexistent,
stable, metastable and stable again. There exists therefore one second
order phase transition and two first order phase transitions in this
range of parameters.

On the other hand, it is likely that the local  minimum energy state
for $(f_0)'_m<f_0<(f_0)'_t$ corresponds to the minimum accessible
energy of the inhomogeneous phase while the local minimum energy 
state for $(f_0)'_t<f_0<(f_0)_c$ corresponds to the minimum accessible
energy of the homogeneous phase. As a result, the curve
$E_m^{(2)}(f_0)$ would not connect the point $((f_0)_m,E_m)\simeq
(0.1075,0.59133)$ as shown in Fig. \ref{f_0-U_2} but rather the point
$((f_0)'_m,E'_m)\simeq (0.1075,0.59473)$ as shown in
Fig. \ref{ground}.

Additional numerical simulations would be necessary to ascertain these
results but they need to be very accurate and we experienced numerical
problems when approaching the minimum energy.

\end{document}